\documentclass[reprint,twocolumn,secnumarabic,amssymb, nobibnotes, aps, pra,superscriptaddress, floatfix]{revtex4-1}

\usepackage{extarrows}
\usepackage{amsmath}    
\usepackage{graphicx}    
\usepackage{verbatim}    
\usepackage{color}           
\usepackage{subfigure}   
\usepackage{braket}
\usepackage{hyperref}    
\usepackage{verbatim}  
\usepackage{here}
\usepackage{mathrsfs}
\usepackage[english]{babel}
\usepackage[utf8x]{inputenc}
\usepackage[T1]{fontenc}
\usepackage{comment}
\usepackage[version=4]{mhchem}

\begin{document}

\title{Satellites in the Ti~1$s$ core level spectra of \ce{SrTiO3} and \ce{TiO2}}

\author{Atsushi~Hariki}
\author{Keisuke~Higashi}
\author{Tatsuya Yamaguchi}
\affiliation{Department of Physics and Electronics, Graduate School of Engineering, Osaka Metropolitan University 1-1 Gakuen-cho, Nakaku, Sakai, Osaka 599-8531, Japan.}

\author{Jiebin~Li}
\author{Curran~Kalha}
\affiliation{Department of Chemistry, University College London, 20 Gordon Street, London WC1H 0AJ, United Kingdom.}%

\author{Manfred~Mascheck}
\affiliation{Scienta Omicron GmbH, Limburger Strasse 75, 65232 Taunusstein, Germany.}

\author{Susanna~K.~Eriksson}
\author{Tomas Wiell}
\affiliation{Scienta Omicron AB, P.O. Box 15120, 750 15 Uppsala, Sweden.}

\author{Frank~M.~F.~de~Groot}
\affiliation{Inorganic Chemistry \& Catalysis, Debye Institute for Nanomaterials Science, Utrecht University, Universiteitsweg 99, 3584 CG, Utrecht, The Netherlands.}

\author{Anna~Regoutz}
\email{a.regoutz@ucl.ac.uk}
\affiliation{Department of Chemistry, University College London, 20 Gordon Street, London WC1H 0AJ, United Kingdom.}%

\date{\today}

\begin{abstract}
Satellites in core level spectra of photoelectron spectroscopy (PES) can provide crucial information on the electronic structure and chemical bonding in materials, particular in transition metal oxides. This paper explores satellites of the Ti~1$s$ and 2$p$ core level spectra of \ce{SrTiO3} and \ce{TiO2}. Conventionally, soft x-ray PES (SXPS) probes the Ti~2$p$ core level; however, it is not ideal to fully capture satellite features due to its inherent spin-orbit-splitting (SOS). Here, hard x-ray PES(HAXPES) provides access to the Ti 1$s$ spectrum instead, which allows us to study intrinsic charge responses upon core-hole creation without the complication from SOS and with favorable intrinsic linewidths. The experimental spectra are theoretically analyzed by two impurity models, including an Anderson impurity model (AIM) built on local density approximation (LDA) and dynamical mean-field theory (DMFT), and a conventional TiO$_6$ cluster model. The theoretical results emphasize the importance of explicit inclusion of higher-order Ti-O charge-transfer processes beyond the nearest-neighboring Ti-O bond to simulate the core level spectra of \ce{SrTiO3} and \ce{TiO2}. The AIM approach with continuous bath orbitals provided by LDA+DMFT represents the experimental spectra well. Crucially, with the aid of the LDA+DMFT method, this paper provides a robust prescription of how to use the computationally cheap cluster model in fitting analyses of core level spectra.
\end{abstract}

\maketitle

\section{Introduction}

Transition metal oxides (TMOs) display a rich variety of functional properties, such as high-temperature superconductivity and colossal magnetoresistance~\cite{imada98,khomskii14}, that have afforded them a high level of interest both fundamentally and in applications. Their properties emerge from atomic Coulomb multiplets embedded into their lattice, forming covalent bonds between transition metal (TM) and ligand orbitals. The nature of their electronic structure and chemical bonding is crucial to explain their functional properties, and photoelectron spectroscopy (PES) has been established as a powerful technique to directly probe both aspects of TMOs. Hard x-ray PES (HAXPES) has provided particularly useful insights regarding the bulk properties of these materials~\cite{BORGATTI201695, Kalha_2021_HAXPES_review}. By going beyond the 2~keV maximum excitation energy of soft x-ray PES (SXPS), HAXPES increases the probing depth significantly enabling the study of the bulk rather than the surface of a material. In the case of \ce{TiO2} for example, the maximum inelastic mean free path, as calculated using the TPP-2M approach implemented in the QUASES software package~\cite{Tanuma1993}, increases from 2.8~nm at the common soft x-ray excitation energy of 1.487~keV (Al~K$\alpha$) to 12.7~nm at the hard x-ray energy of 9.252~keV used in this paper.\par

In the case of 3$d$ TMOs, the most widely studied core level using PES is the 2$p$ state, which delivers rich information on their electronic structure ranging from metal-insulator transitions to magnetic and orbital ordering~\cite{Horiba04,veenendaal06,Hariki17,Taguchi08,Eguchi08,Obara10,Kamakura04,Hariki13b,Taguchi05,Chang18}. This is possible as the electronic response to the local perturbation (i.e., the creation of a core hole) gives rise to specific spectral features due to charge-transfer screening from surrounding ions via the underlying chemical bonding~\cite{veenendaal06,veenendaal93,Hariki17,Ghiasi2019}. However, the analysis and interpretation of the already complex satellite structures present in the 2$p$ state are further complicated by the presence of spin-orbit-splitting (SOS) effects leading to overlap of spectral features within the core spectral range. Here, HAXPES brings an additional advantage over SXPS in providing access to deeper core states, which can be advantageous for spectral analysis due to absence of SOS, favorable intrinsic linewidths, reduction of overlap with other spectral features including core and Auger lines, and higher photo-ionization cross-sections~\cite{Miedema2015,Young2015,Woicik2015,Rubio2018,Ghiasi2019,Berens2020,Siol2020,Kalha2022}. This has already been exploited in the case of 3$d$ TMOs, by accessing their 1$s$ core states using HAXPES, particularly for the late TMOs~\cite{Kamakura04, Calandra_2012, Miedema2015, Ghiasi2019, Thien_2020}. 

Due to the complexity of the spectra, theoretical approaches to aid their interpretation are crucial. 
Since the core hole does not move and it couples exclusively to the localized $d$ electrons on the same TM site,
an impurity model representing the excited ion is a good starting point for modeling core level spectra. For 3$d$ TMOs, the MO$_6$ cluster model is widely employed~\cite{groot_kotani,Bocquet92}. It includes the x-ray excited metal and the surrounding ligands, thus implementing metal-oxygen (M--O) hybridization on the nearest-neighboring bond. Though the cluster model serves as a convenient platform for simulating spectra, it can suffer from a number of limitations:~(1) hybridization between long-distance M--O and M--M bonds is lacking in the cluster model, which may be relevant for a charge response to the core hole, and (2) electronic configurations accompanied by higher-order electron exchange with the ligands, which are usually discarded to make the computation feasible, may affect the simulated spectra. Especially for highly covalent early TMOs, these limitations modify the model parameters during fitting of experimental data as well as impact the interpretation of spectral features.\par 

This paper explores the Ti 1$s$ and 2$p$ core level spectra of \ce{SrTiO3} and \ce{TiO2}, prototype titanium oxides, using both experiment and theory. HAXPES is used to enable access to their Ti~1$s$ spectra and bulk information on their electronic structure. Experimental results are fitted by two theoretical impurity models:~(1) a TiO$_6$ cluster model and (2) an Anderson impurity model (AIM) built on local density approximation (LDA) combined with dynamical mean-field theory (DMFT)~\cite{Hariki17,Ghiasi2019,kotliar06,georges96}. The latter can be viewed as an extension of the cluster model to incorporate hybridization among long-distance bonds, while retaining the impurity model description. This is achieved by replacing the ligand orbitals in the cluster model by the continuous bath provided by LDA+DMFT, which represents an electron exchange with the distant ions in the periodic lattice. LDA+DMFT AIM has already been employed successfully to identify spectral features due to long-distance charge transfer in 2$p$ and 1$s$ x-ray photoelectron spectroscopy (XPS) spectra of mid- and late TMOs~\cite{Hariki17,Ghiasi2019,hariki16,Higashi21}. Here, this method is expanded to titanium oxides to examine the validity of the impurity model description for core level spectra of early TMOs, where covalency plays a crucial role.\par

\section{Method}
\subsection{Experimental method}
Two single crystals of rutile \ce{TiO2} (110) and \ce{SrTiO3} (100) were used for the measurements. The \ce{SrTiO3} crystal was 1\% doped with Nb to increase its conductivity. Both crystals were purchased from CRYSTAL, and no further sample preparation was needed. HAXPES measurements were performed on a Scienta Omicron HAXPES Lab system~\cite{Regoutz2018,Hashimoto2021}. This spectrometer uses a monochromated Ga~K$\alpha$ x-ray source, giving an excitation energy of 9.252~keV, and a Scienta Omicron EW4000 hemispherical electron energy analyzer to collect the excited photoelectrons. A pass energy of 200~eV, grazing incidence geometry ($< 3^\circ$), and near-normal emission geometry were used for all measurements. The total energy resolution of this setup is 485~meV (16/84\% width of the Au $E_F$). More details about the experimental setup can be found in a previous publication~\cite{Regoutz2018}. Complementary SXPS measurements were performed on a Thermo Scientific K-Alpha XPS system, which uses a monochromated Al~K$\alpha$ x-ray source ($h\nu$~=~1.487~keV). Measurements were conducted with a 400~$\mu$m spot size and a flood gun was used for charge compensation. Pass energies of 20 and 15~eV were used for core and valence spectra, respectively. The total energy resolution at 15~eV is 420~meV (16/84\% width of the Au $E_F$).\par

\subsection{Computational method} 
The Ti~1$s$ HAXPES simulations start with a standard LDA+DMFT calculation~\cite{georges96,kotliar06,kunes09}. The LDA bands obtained for the experimental structures of the studied compounds are projected onto a tight-binding model spanning Ti~3$d$ and O~2$p$ orbitals~\cite{wien2k,wien2wannier,wannier90}.
The tight-binding model was augmented with a local electron-electron interaction within the Ti~$3d$ shell, defined by Hubbard $U$ and Hund's $J$ parameters with values of $(U,J)$=(4.78~eV, 0.64~eV), which are chosen by consulting with previous density functional theory (DFT)-based and spectroscopy studies for titanates including \ce{SrTiO3} and \ce{TiO2}~\cite{Okada94,Tanaka94,bocquet96,Lechermann17,Pavarini04,Okamoto06}.

The present LDA+DMFT implementation follows Refs.~\cite{Hariki17,Ghiasi2019,Higashi21}. The strong-coupling continuous-time quantum Monte Carlo impurity solver~\cite{werner06,boehnke11,hafermann12,Hariki15} is used to compute the self-energies $\Sigma(i\omega_n)$ of Ti~$3d$ electrons from the AIM. In the LDA+DMFT scheme, the bare energies of these $d$ states are obtained from the LDA values by subtracting the so-called double-counting correction $\mu_{\rm dc}$, which accounts for the effect of the interaction already present in the LDA result~\cite{karolak10,kotliar06}. Its appropriate value is determined by comparing the LDA+DMFT result with the experimental PES and the band gap, as discussed below.
All calculations are performed at $T=300$~K. After converging the DMFT calculation, analytically continued $\Sigma(\varepsilon)$ in the real-frequency domain is obtained using the maximum entropy method~\cite{jarrell96}. It is then used to calculate the real-frequency one-particle spectral densities and hybridization densities $V^2(\varepsilon)$. The latter represents the exchange of electrons between the Ti ion and the rest of the crystal.

Ti~1$s$ HAXPES spectra were computed from AIM with the DMFT hybridization densities $V^2(\varepsilon)$. AIM includes the Ti~1$s$ core orbital and its Coulomb interaction with 3$d$ electrons explicitly. The Coulomb interaction parameter between the 1$s$ hole and Ti~3$d$ electrons $U_{dc}$ is set to
$U_{dc} = 1.3 × U_{dd}$, where $U_{dd}$ is the configuration averaged Coulomb interaction between Ti~3$d$ electrons, and the value is $U_{dd}=4.5$~eV for the used Hubbard $U$ and Hund's $J$ values~\cite{Ghiasi2019}. 
This is a well-established empirical rule in simulating core level XPS of 3$d$ TMOs~\cite{Hariki17}. A configuration-interaction (CI) method for computing the HAXPES intensities is employed, for details see Refs.~\cite{Hariki17,Winder20}. The CI scheme, which is widely used in computing the core level spectra using an impurity-based model, generates basis configurations by a sequential exchange of electrons between the impurity site and the bath (representing the rest of the crystal) starting with a normal-valence configuration, i.e.,~$|d^0\rangle$ for tetravalent Ti systems. The initial state $|g\rangle$ is described by a linear combination of the configurations as 

\begin{eqnarray}
  |g\rangle&=&
  |d^0\rangle + |d^1 \underline{L} \rangle + |d^2 \underline{L}^2 \rangle + |d^3 \underline{L}^3 \rangle+|d^4 \underline{L}^4 \rangle +
   \cdots. \notag
\end{eqnarray}

Here, $\underline{L}$ denotes a hole in the valence orbitals of the bath, and thus, $|d^n \underline{L}^m \rangle$ represents an electronic configuration with $n$ $d$-electrons in the impurity Ti site and $m$ holes in the valence bands. Ti~1$s$ HAXPES final states are then described by the configurations above plus a core hole in the Ti~1$s$ shell. Spectra calculated by the conventional TiO$_6$ cluster model are also presented.
Though the cluster model implements the same intra-atomic interactions as the LDA+DMFT AIM, the valence orbitals consist of only the 2$p$ orbitals on the nearest-neighboring ligands. The hybridization strength between the Ti~3$d$ and O~2$p$ orbitals of the cluster model is extracted from the tight-binding model construed above, and the values are provided below. Though the CI scheme provides a systematic way to include the hybridization effect starting from the isolated atomic limit ($|d^0\rangle$), care may need to be taken for the cutoff in the basis expansion above, which will be discussed below. 

In previous studies for Ti~2$p$, 3$s$, and 3$p$ core level XPS spectra of Ti oxides using the TiO$_6$ cluster model~\cite{Okada94,Okada93}, three electronic configurations ($|d^0\rangle$, $|d^1 \underline{L} \rangle$ and $|d^2 \underline{L}^2 \rangle$) are considered in the spectral analysis.

To enable a direct comparison between the theoretically obtained projected density of states (PDOS) and the experimental valence band spectra, the PDOS results were broadened and photo-ionization cross-section corrected. Broadening and cross-sections were chosen to match the Al~K$\alpha$ SXPS measurements, including a Gaussian broadening of 420~meV commensurate with the total energy resolution of the experiment and Scofield cross-sections~\cite{Scofield76, Kalha2020}.

\section{Results and Discussion}

HAXPES survey spectra of the \ce{SrTiO3} and \ce{TiO2} single crystals show only the expected core and Auger lines of the oxides with no contaminants detectable (see Fig.~S1 in the Supplemental Material). In addition to the Ti core state and valence band spectra, which will be discussed in detail in the following, the O~1$s$ for both samples as well as the Sr~2$s$, 2$p_{3/2}$ and 3$d$ core state spectra of \ce{SrTiO3} were collected for completeness (see Fig.~S2 in the Supplemental Material). Figure~\ref{fig:Ti_CLs}(a) shows the Ti~1$s$ and 2$p$ HAXPES core level spectra of both crystals. The Ti~2$p$ spectra show the 2$p_{3/2}$ and 2$p_{1/2}$ components with a SOS of 5.7~eV. The HAXPES and SXPS Ti~2$p$ spectra are comparable (see Fig.~S3 in the Supplemental Material). In contrast, the Ti~1$s$ does not exhibit SOS and therefore is advantageous for the identification of satellite features as the overall spectral shape is simplified. In comparison to the 2$s$ and 3$s$ core levels, which also do not exhibit SOS, the 1$s$ line has the lowest intrinsic linewidth (0.89~eV compared with 3.9~eV for 2$s$ and 2.1~eV for 3$s$)~\cite{Campbell2001}, aiding the identification of low-energy satellite features.

\begin{figure}
    \includegraphics[width=0.99\columnwidth]{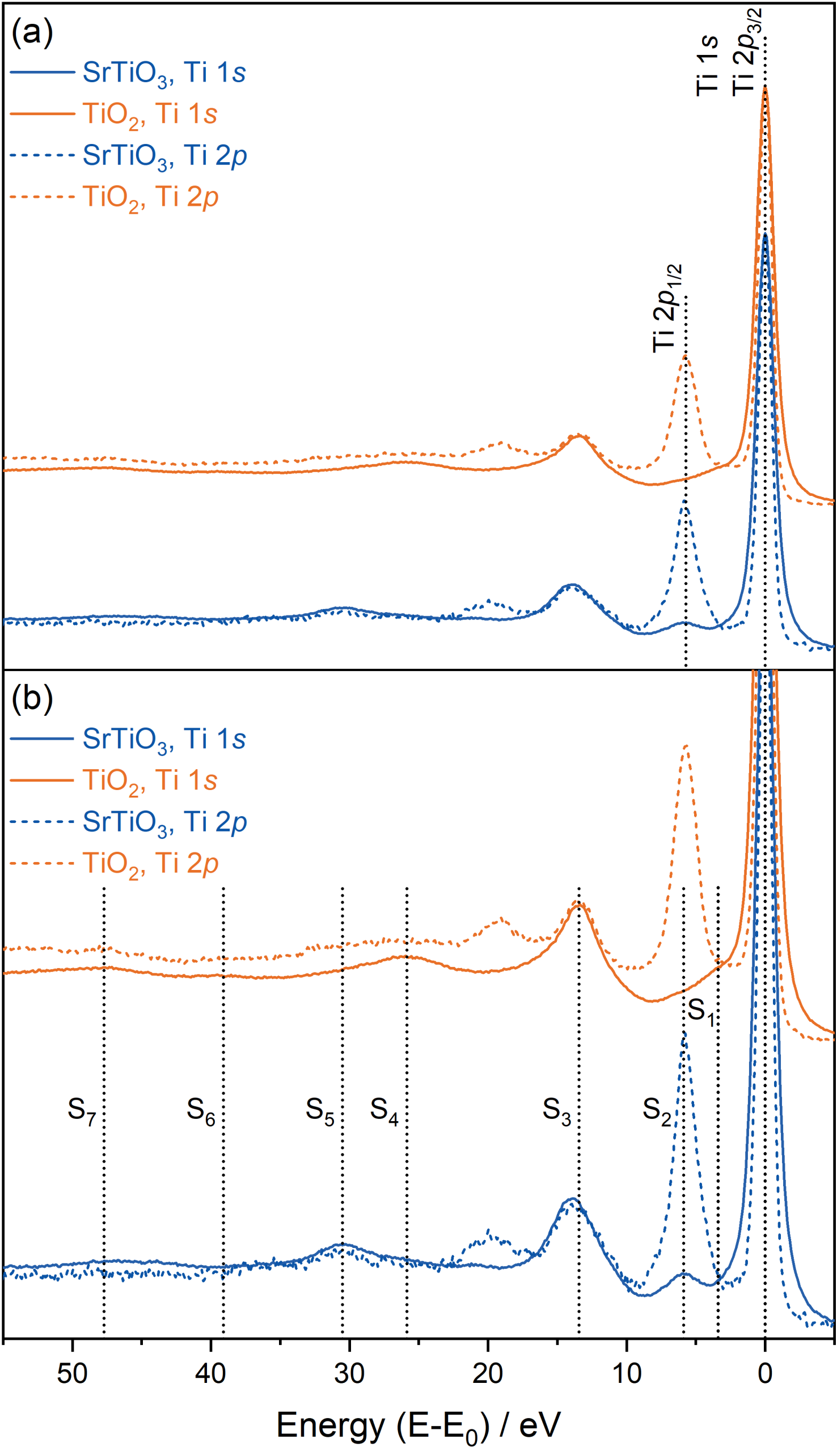}
    \caption{Hard x-ray photoelectron spectroscopy (HAXPES) Ti~1$s$ and 2$p$ spectra of \ce{SrTiO3} and \ce{TiO2}, including (a) complete spectra and (b) magnified view of the satellite structure. All spectra are aligned to the main peak (1$s$ and 2$p_{3/2}$) at 0~eV and a relative energy scale shown. The guidance lines in (b) are taken from the Ti~1$s$ spectra at the satellite positions for \ce{TiO2} except for $S_2$, which is only clearly observed in \ce{SrTiO3}.}
    \label{fig:Ti_CLs}
\end{figure}

Figure~\ref{fig:Ti_CLs}(b) shows an expanded view of the experimentally observed satellite features $S_1$-$S_7$ with position guidelines based on the Ti~1$s$ spectra for \ce{TiO2} except for $S_2$, which is only clearly observed in \ce{SrTiO3}. The energy positions of the satellites are summarized in Table~\ref{tab:sat}. In the 2$p$ spectra, satellite features appear twice due to SOS at a separation of 5.7~eV, e.g., $S_3$ at 13.4 and the spin-orbit-split peak at 19.2~eV. While authors of a multitude of studies have reported on the satellites observed in the Ti~$2p$ spectra, authors of only two previous studies report satellites in the Ti~1$s$ spectra of \ce{SrTiO3} and \ce{TiO2}, noting satellite features at 5, 13, and 26~eV~\cite{Moslemzadeh2006,Woicik2015}. These values are in good agreement with the data presented here but do not include a discussion of the additional spectral features observed in this paper. Features $S_4$ and $S_5$ are associated with the Ti core level excitation, as will be discussed below.\par

\begin{table}
    \centering
    \caption{Experimentally observed satellite positions relative to the main peaks (1$s$ and 2$p_{3/2}$) at 0~eV. Features that cannot be clearly identified in the experimental data are denoted as not detectable (n.d.).}
    {\tabcolsep =0.2cm
    \begin{tabular}{lrrrrrrr}
    \hline \hline
         & $S_1$ & $S_2$ & $S_3$ & $S_4$ & $S_5$ &$S_6$ & $S_7$ \\
    \hline
     \ce{TiO2}  & 3.4 & 5.7 & 13.4 & 25.9 & 30.5 & 39.1 & 47.7 \\
     \ce{SrTiO3}   & n.d. & 5.9 & 13.9 & 25.8 & 30.5 & n.d. & 46.6\\
    \hline \hline
    \end{tabular}
    }
    \label{tab:sat}
\end{table}

Authors of previous cluster model studies for Ti 2$p$ XPS of \ce{TiO2}~\cite{Okada94,Okada93} explained that the two features (i) and (iii) at 0~eV and 13~eV correspond to bonding- and antibonding-split final states, respectively.
The large energy splitting of the two final states is due to a strong Ti--O covalent bonding, i.e.,~a large hybridization between the $|d^0\rangle$ and charge-transferred $|d^1\underline{L}\rangle$ electronic configurations leads to a formation of well-defined bonding and antibonding states.

\begin{figure}
    \includegraphics[width=0.99\columnwidth]{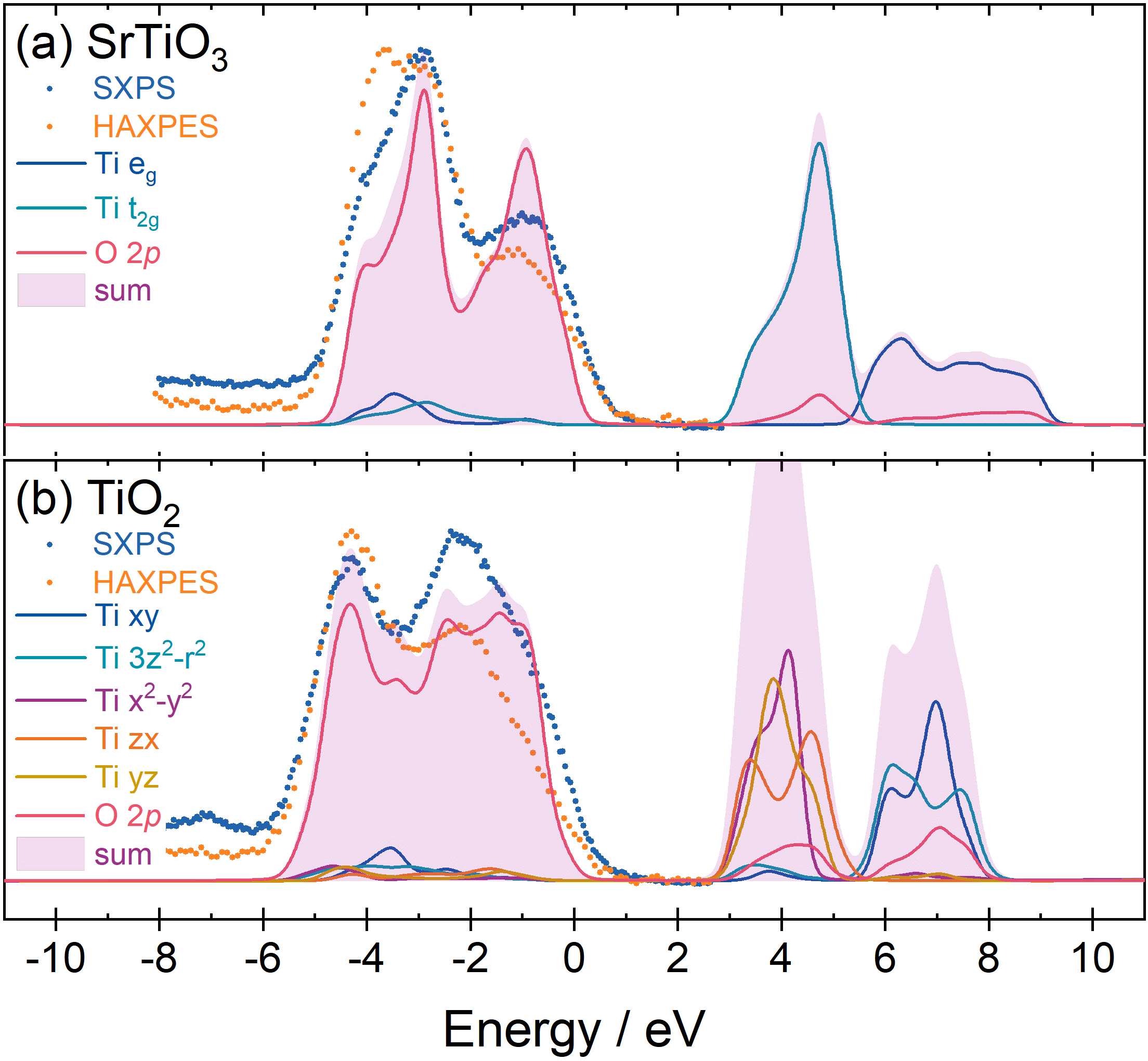}
    \caption{Soft x-ray photoelectron spectroscopy (SXPS) and hard x-ray photoelectron spectroscopy (HAXPES) valence spectra and broadened, one-electron photo-ionization cross-section weighted local density approximation and dynamical mean-field theory (LDA+DMFT) projected density of states (PDOS) for (a) \ce{SrTiO3} and (b) \ce{TiO2}. The sum of the individual PDOS contributions is also shown. The broadening and cross-section corrections were chosen to match the SXPS experimental setup. In the LDA+DMFT results, $\mu_{\rm dc}=3.0$~eV was used. Experimental data were aligned to the O~2\textit{p}-dominated features at the bottom of the valence band.}
    \label{fig:VB_PDOS_comp}
\end{figure}

Before examining the multiple satellite features observed in the Ti~1$s$ HAXPES spectra and discussing appropriate theoretical models of the core level excitations in the studied Ti compounds, the electronic structure calculations, which form the basis for the core level spectral calculations using LDA+DMFT AIM, are discussed. To validate the computational parameter, i.e.,~double-counting correction value $\mu_{\rm dc}$ used in the LDA+DMFT self-consistent calculation, PDOS of both \ce{SrTiO3} and \ce{TiO2} are compared with SXPS and HAXPES valence spectra in Fig.~\ref{fig:VB_PDOS_comp}. A good agreement in the overall shape and relative energy positions of features of the valence band states is found between theory and SXPS spectra. Practically, $\mu_{\rm dc}$ renormalizes the energy levels of the metal~$3d$ to the O~2$p$ orbitals~\cite{karolak10,Higashi21,Haule15}. Thus, for a band insulator with a gap between empty metal 3$d$ and filled O~2$p$ bands, which is the case for \ce{SrTiO3} and \ce{TiO2}, $\mu_{\rm dc}$ can be estimated by reproducing the experimental band gap. Here, $\mu_{\rm dc}=3.0$~eV yields good agreement to previously reported experimental bandgap ($\sim 3$~eV) or inverse PES data~\cite{Tezuka94,Ohtomo04}. The $\mu_{\rm dc}$ determination can be found in Part B of the Supplemental Material (Figs.~S4 and S5).

The direct comparison of the theoretical PDOS with the HAXPES spectra illustrates the influence of the energy-dependent photo-ionization cross-sections. The relative increase in Ti~$s$ state cross-sections at higher x-ray photon energy leads to an increase in overall intensity at the bottom of the valence band. This ability to enhance \textit{s} contributions represents another key advantage of HAXPES, which has been previously exploited to probe the valence band orbital character of other metal oxide systems~\cite{Payne_2007, Mudd_2014, Takegami_2019}. As the Ti~$s$ and $p$ as well as the O $s$ states are not explicitly included in the LDA+DMFT calculations, the theory was corrected for the SXPS setup as, due to photo-ionization cross-section effects, the contributions from Ti~$d$ and O~$p$ states dominate at lower photon energies. The unbroadened, uncorrected theoretical PDOS results can be found in Fig.~S6 in the Supplemental Material.

\begin{figure}
    \includegraphics[width=0.99\columnwidth]{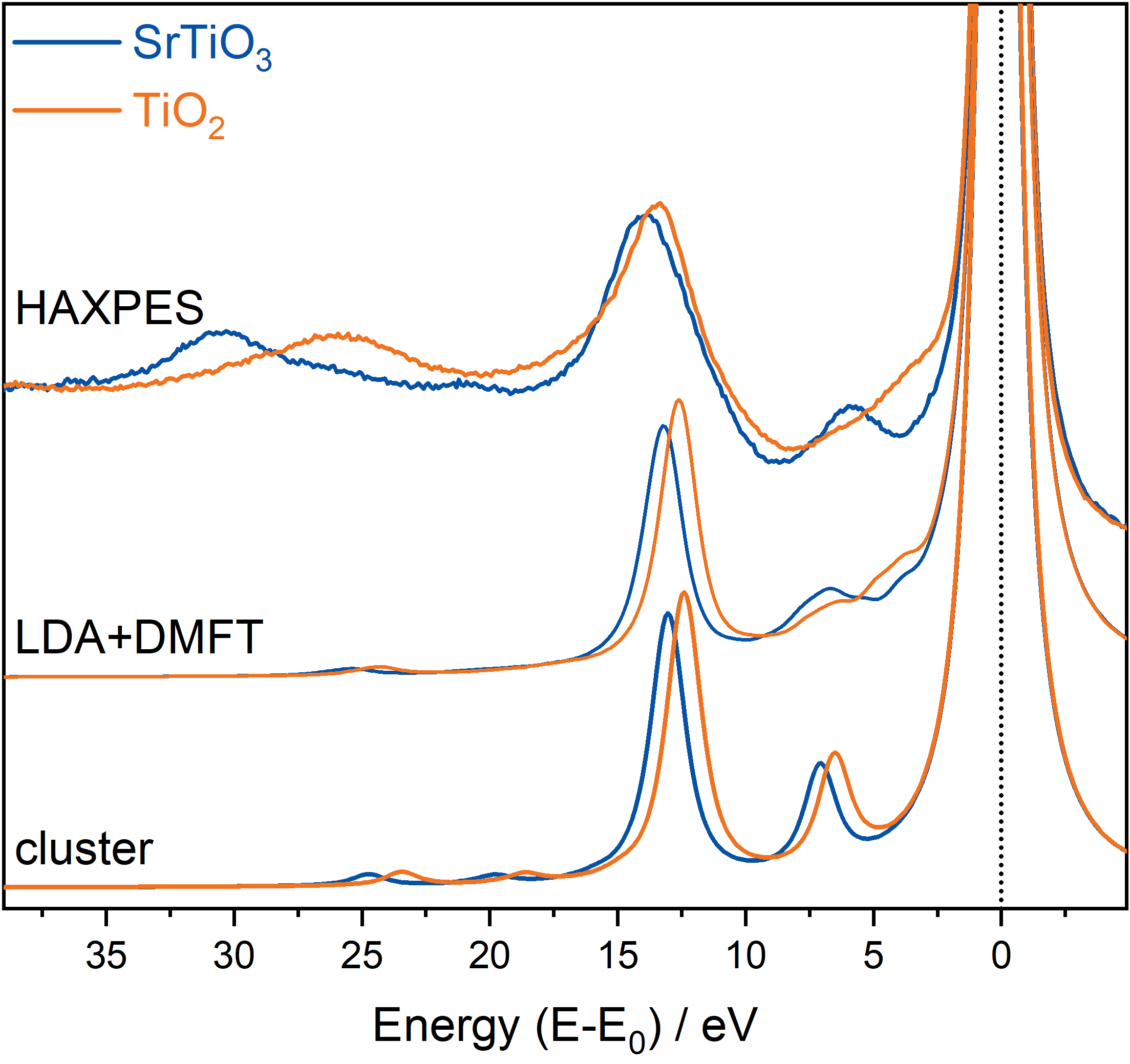}
    \caption{Ti~1$s$ spectra calculated by the cluster and local density approximation and dynamical mean-field theory (LDA+DMFT) Anderson impurity model (AIM) methods and from hard x-ray photoelectron spectroscopy (HAXPES) experiments for \ce{SrTiO3} and \ce{TiO2}. All spectra are aligned to the 1$s$ main peak at 0~eV, and a relative energy scale is shown.}
    \label{fig:Ti1s_exptheor}
\end{figure}

Building upon the electronic structure model, Ti~1$s$ core level spectra were computed using the LDA+DMFT AIM for \ce{SrTiO3} and \ce{TiO2}. Both experimental HAXPES and simulated Ti~1$s$ spectra are shown in Fig.~\ref{fig:Ti1s_exptheor}. The simulated spectra can reproduce the characteristic satellite features up to 30~eV above the main peak (including satellites $S_1$-$S_4$ in the experimental spectra) in both compounds, with energy positions, relative intensities and spectral shapes captured. The relative energy shift of the most intense feature $S_3$ is clear in both theory and experiment although theory underestimates the width of this feature. The low-energy satellites $S_1$ and $S_2$ are particularly well matched between experiment and theory for both \ce{TiO2} and \ce{SrTiO3}, indicating that they are indeed intrinsic to the materials and have been missed in previous experiments focusing on Ti~$2p$ spectra. Theory can also reproduce satellite $S_4$ at just below 26~eV in the experiment.\par

\begin{figure}
    \includegraphics[width=0.99\columnwidth]{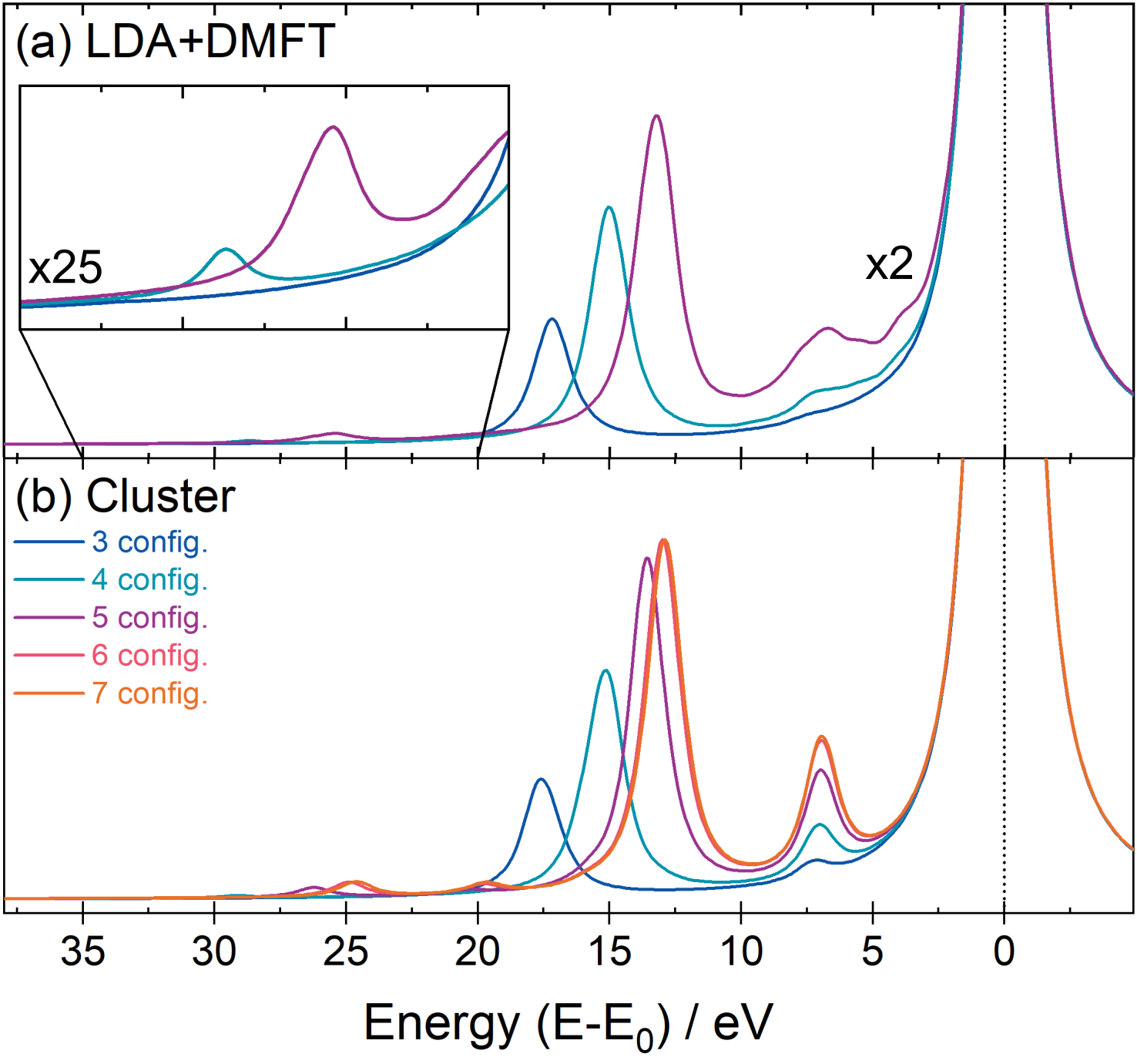}
    \caption{\ce{SrTiO3} Ti~1$s$ hard x-ray photoelectron spectroscopy (HAXPES) spectra calculated by (a) local density approximation and dynamical mean-field theory (LDA+DMFT) Anderson impurity model (AIM) and (b) TiO$_6$ cluster model with varying number of electronic configurations (config.) included in the spectral simulation:~three config. ($|d^0\rangle+ |d^1\underline{L}\rangle + |d^2\underline{L}^2\rangle$),
    four config. ($|d^0\rangle+|d^1\underline{L}\rangle+|d^2\underline{L}^2\rangle+|d^3\underline{L}^3\rangle$), five config. ($|d^0\rangle+|d^1\underline{L}\rangle+|d^2\underline{L}^2\rangle+
    |d^3\underline{L}^3\rangle+|d^4\underline{L}^4\rangle$), six config. ($|d^0\rangle+|d^1\underline{L}\rangle+|d^2\underline{L}^2\rangle+
    |d^3\underline{L}^3\rangle+|d^4\underline{L}^4\rangle+|d^5\underline{L}^5\rangle$), 
    and 
    seven config. ($|d^0\rangle+|d^1\underline{L}\rangle+|d^2\underline{L}^2\rangle+
    |d^3\underline{L}^3\rangle+|d^4\underline{L}^4\rangle+|d^5\underline{L}^5\rangle+|d^6\underline{L}^6\rangle$). The inset in (a) shows a magnified view of the high-energy region of the spectra. All spectra are aligned to the main peak at 0~eV, and a relative energy scale is shown.}
    \label{fig:Ti1s_theor}
\end{figure}

Given the good agreement between the theoretical calculations and the experimental data, the character of the spectral features can be identified based on further exploration of the theoretical parameters. Figure~\ref{fig:Ti1s_theor}(a) shows the simulated Ti~1$s$ spectra of \ce{SrTiO3} calculated with varying number of electronic configurations included. Satellite $S_3$, and to a much subtler degree $S_2$, moves to lower energy relative to the main peak when higher configurations ($|d^2 \underline{L}^2 \rangle, |d^3\underline{L}^3 \rangle$) are included. This behavior can be explained as the energies of these configurations with multiple electrons in the Ti~3$d$ shell are rather high ($> 25$~eV) due to the energy cost from the onsite $d$--$d$ Coulomb repulsion, see Appendix~B for the estimated values.  However, the higher configurations are coupled to low-lying configurations ($|d^0\rangle, |d^1\underline{L}^1\rangle$) via the strong Ti--O covalent bonding, which renormalizes the entire spectrum, moving features $S_3$ and $S_2$ to lower energies. This indicates that, when implementing an impurity model analysis for core level PES of highly covalent TMOs, care must taken regarding the number of electronic configurations included in the numerical simulation.\par

Figure~\ref{fig:Ti1s_theor} further explores a range of cluster model parameters and their validity for \ce{SrTiO3}. In the TiO$_6$ cluster model, the hopping parameter $V$ on the nearest-neighboring Ti--O bonds is taken from the tight-binding model constructed from the LDA bands. It gives estimates of $V_{e_g}=4.01$~eV and $V_{t_{2g}}=-2.33$~eV for the Ti~$e_g$ and $t_{2g}$ orbitals, respectively. These values are consistent with a previous DFT-based estimate by Haverkort \textit{et al.}~\cite{Haverkort12}. 
The parameter values of the present cluster model are summarized in Appendix~B.
The cluster model spectra including the basis configuration dependence, see Fig.~\ref{fig:Ti1s_theor}(a), resemble the LDA+DMFT spectra in Fig.~\ref{fig:Ti1s_theor}(b). Thus, the cluster model description works reasonably well for the Ti core level of \ce{SrTiO3}. However, the low-energy satellite features $S_1$ and $S_2$ in the cluster model are much sharper than in the LDA+DMFT AIM. This difference suggests that these satellites are related to the band structure since the LDA+DMFT description explicitly considers the O~2$p$ bands, whereas only O~2$p$ discrete levels on the nearest-neighboring ligands are included in the cluster model. The cluster model result also shows that the Ti~1$s$ spectra are well converged by including up to five configurations in the basis expansion. In earlier studies using the cluster model implementing up to three configurations ($|d^0\rangle+ |d^1\underline{L}\rangle + |d^2\underline{L}^2\rangle$)~\cite{Okada94}, the hopping parameter derived from a fitting analysis of experimental Ti~2$p$ XPS data is $\sim$25\% smaller than the DFT-based estimate above.
This suggests that the higher electronic configurations must be included in the fitting analysis of Ti core level spectra of Ti oxides.\par

\begin{figure}
    \includegraphics[width=0.99\columnwidth]{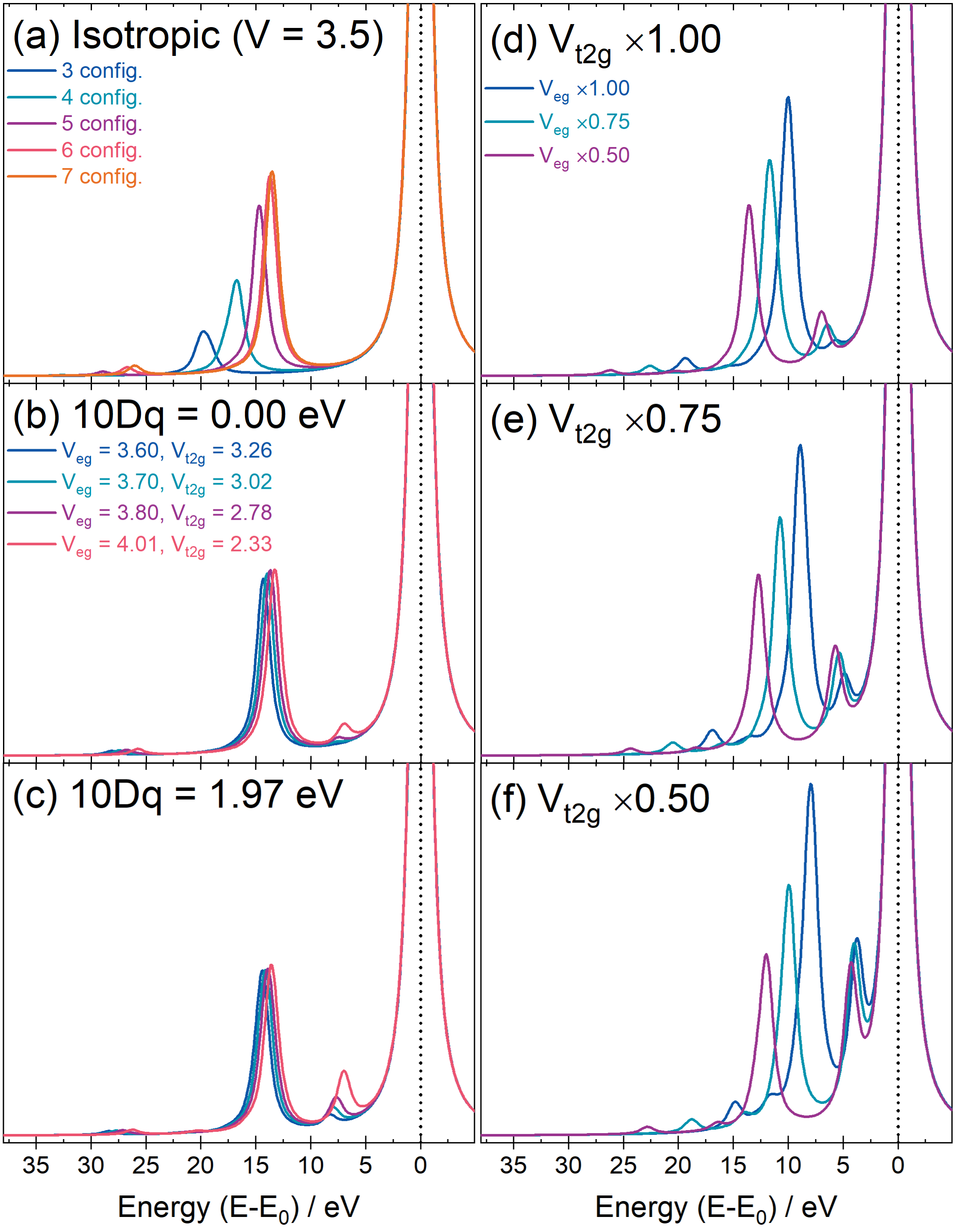}
    \caption{Cluster model 1$s$ spectra simulating \ce{SrTiO3}. (a) Isotropic cluster model spectra computed with different configuration basis, see caption of Fig.~\ref{fig:Ti1s_theor} for details. (b)--(c) Cluster model spectra computed for selected hopping parameters with $10Dq=0.00$ and $1.97$~eV [the local density approximation (LDA) value], respectively. (d)--(f) Cluster model spectra computed by varying the $V_{t_{2g}}$ and $V_{e_{g}}$ values independently. The five-configuration basis is employed in panels (b)--(f). All spectra are aligned to the main peak at 0~eV and a relative energy scale is shown.}
    \label{fig:cluster}
\end{figure}

The Ti~1$s$ spectra computed by the isotropic cluster model, where $V_{e_g}=V_{t_{2g}}=3.5$~eV and the crystal-field splitting (10$Dq$) is set to zero, (see Fig.~\ref{fig:cluster}(a)), show satellite $S_3$ clearly, and its configuration dependence resembles the realistic cluster model in Fig.~\ref{fig:Ti1s_theor}(b). However, the low-energy satellites $S_1$ and $S_2$ are not reproduced in the isotropic cluster model result. By switching the anisotropy in the hybridization on (i.e.,~$V_{e_g}\neq V_{t_{2g}}$), a satellite feature does appear albeit at a slightly higher energy of 7~eV than experiment and LDA+DMFT. It is worth noting that the satellite does not split from the satellite $S_3$. This behavior suggests that this satellite feature is related to a nonbonding state for the bonding state (main line) and the antibonding state forming satellite $S_3$. The simulation results allow modeling of the low-energy bonding properties of the studied Ti oxides by a very simple model provided in Appendix~A.
The crystal field splitting enhances the intensity of the low-energy satellite feature, as can be seen in Fig.~\ref{fig:cluster}(c).
To emphasize the orbital character of the two satellites, Figs.~\ref{fig:cluster}(d)--~\ref{fig:cluster}(f) show the cluster model spectra calculated with rescaled hopping parameters for $e_g$ and $t_{2g}$ orbitals. Here, the five-configuration basis expansion is employed in the spectral evaluation. The $V_{t_{2g}}$ hopping mainly modifies the binding energy of the low-energy satellite, whereas $V_{e_{g}}$ hopping controls that of satellite $S_3$. 

Finally, a clear material dependence in the satellite $S_3$ composed of the antibonding-split final states is observed. 
Since the bonding and antibonding splitting is determined largely by the hybridization on the nearest-neighboring Ti--O bond, the $S_3$ satellite of the cluster model is almost on top of the one in the LDA+DMFT AIM for both compounds, see Fig.~\ref{fig:Ti1s_exptheor}.
This is in clear contrast to satellite $S_2$, as discussed in Figs.~\ref{fig:Ti1s_exptheor} and \ref{fig:Ti1s_theor}. The $S_3$ satellite of \ce{TiO2} is $\sim$$0.5$~eV shallower than that of \ce{SrTiO3}, indicating a weaker Ti--O bonding in the former. This can be quantified by the effective hybridization strength $V_{\rm eff}$ that represents the coupling between the formal valence configuration $|d^0\rangle$ and the charge-transferred one $|d^1\underline{L}\rangle$~\cite{Okada94,Ghiasi2019}. The effective hybridisation $V_{\rm eff}$ is defined as $\sqrt{4V^2_{e_g}+6V^2_{t_{2g}}}$ for \ce{SrTiO3} and $\sqrt{2V^2_{B_{1g}}+2V^2_{Ag}+2V^2_{Ag'}+2V^2_{B_{2g}}+2V^2_{B_{3g}}}$ for \ce{TiO2}, which amounts to 9.84~eV and 9.49~eV, respectively, from the hopping amplitudes in the LDA result (see Appendix~B). These estimates support a weaker Ti--O bonding in \ce{TiO2} than in \ce{SrTiO3} and emphasize a close relation between the satellite $S_3$ and the Ti--O bonding strength. 

\section{Conclusion}

PES is widely used to probe chemical environments and bonding as well as the electronic structure of TMOs. This paper showcases the usefulness of collecting deeper core level spectra in favor of the more commonly explored 2$p$ states, using the example of Ti~1$s$ spectra of \ce{SrTiO3} and \ce{TiO2}. The lack of SOS and favorable intrinsic linewidths lead to the observation of satellite features not observed previously. The presented theoretical approaches based on LDA+DMFT as well as a conventional cluster model provide a good description of the experimental spectra. The comparison emphasizes the crucial importance of explicitly including higher-order Ti-O charge-transfer processes beyond the nearest-neighboring Ti-O bonds. Finally, this paper confirms that the presented theoretical approaches can provide a successful description of early TMOs, where covalency plays a central role, promising wider applicability to the many technologically crucial materials in this family of compounds.\par 

\begin{acknowledgments}

A.H. was supported by JSPS KAKENHI Grant No. 21K13884 and 21H01003. C.K. acknowledges the support from the Department of Chemistry, UCL. A.R. acknowledges the support from the Analytical Chemistry Trust Fund for her CAMS-UK Fellowship and from Imperial College London for her Imperial College Research Fellowship. The authors would like to thank T. Uozumi for valued discussions.

\end{acknowledgments}

\appendix

\section{Toy model for the low-energy satellites}

Here, a simple toy model for the low-energy satellites in Fig.~\ref{fig:cluster} is proposed. The model consists of three levels labeled as $|0\rangle, |e\rangle$, and $|t\rangle$. Here, $|0\rangle$ represents an ionic tetravalent Ti state, i.e.,~it corresponds to the $|d^0\rangle$ configuration in the AIM or cluster model description. Also, $|e\rangle$ ($|t\rangle$) simulate states with an $e_{g}$ ($t_{2g}$) Ti $3d$ electron and a hole on ligands in the $|d^1\underline{L}\rangle$ configuration. Considering the matrix elements by the charge transfer between the Ti site and ligands, the low-energy excitations of the studied Ti compounds can be modeled by the $3\times3$ Hamiltonian with the basis order \{$|0\rangle,|e\rangle, |t\rangle$\}:

\begin{equation}
H=
\begin{pmatrix}
  0   & t_e & t_t   \\
  t_e & e_e & 0     \\
  t_t & 0   & e_t   \notag
\end{pmatrix},
\end{equation}

where $e_e$ (and $e_t$) account for the charge-transfer energy and the crystal field splitting; thus $e_e\neq e_t$ in reality. The $v_e$ and $v_t$ are the hopping amplitude for the $e_g$ and $t_{2g}$ orbitals with the ligand orbitals, respectively. Next, by applying the hopping term in the Hamiltonian to the $|0\rangle$ state, a state $| b \rangle$, and then an orthogonal state $|n \rangle$ are obtained as

\begin{eqnarray}
|b\rangle = \dfrac{1}{\sqrt{t_e^2+t_t^2}}\left( t_e|e\rangle+t_t|t\rangle \right) \notag \\
\textrm{and}\;|n\rangle = \dfrac{1}{\sqrt{t_e^2+t_t^2}}\left( t_t|e\rangle-t_e|t\rangle \right) \notag.
\end{eqnarray}

Representing the Hamiltonian with the \{$|0\rangle,|b\rangle, |n\rangle$\} basis set,

\begin{equation}
H=
\begin{pmatrix}
  0   & \sqrt{t_e^2+t_t^2} & 0  \\
  \sqrt{t_e^2+t_t^2} & \dfrac{e_et_e^2+e_t t_t^2}{t_e^2+t_t^2} & \dfrac{(e_e-e_t)t_et_t}{t_e^2+t_t^2}   \\
  0 & \dfrac{(e_e-e_t)t_et_t}{t_e^2+t_t^2}   & \dfrac{e_e t_t^2+e_t t_e^2}{t_e^2+t_t^2}      \notag
\end{pmatrix}.
\end{equation}

By the above construction, the $|0\rangle$ and $|n\rangle$ states do not couple. Note that modeling the 1$s$ XPS final states shifts the energies $e_e$ and $e_t$ by the core-hole potential $U_{dc}$ due to the presence of the 1$s$ core hole; thus the structure of the Hamiltonian above does not change. 

\begin{figure}
    \includegraphics[width=0.99\columnwidth]{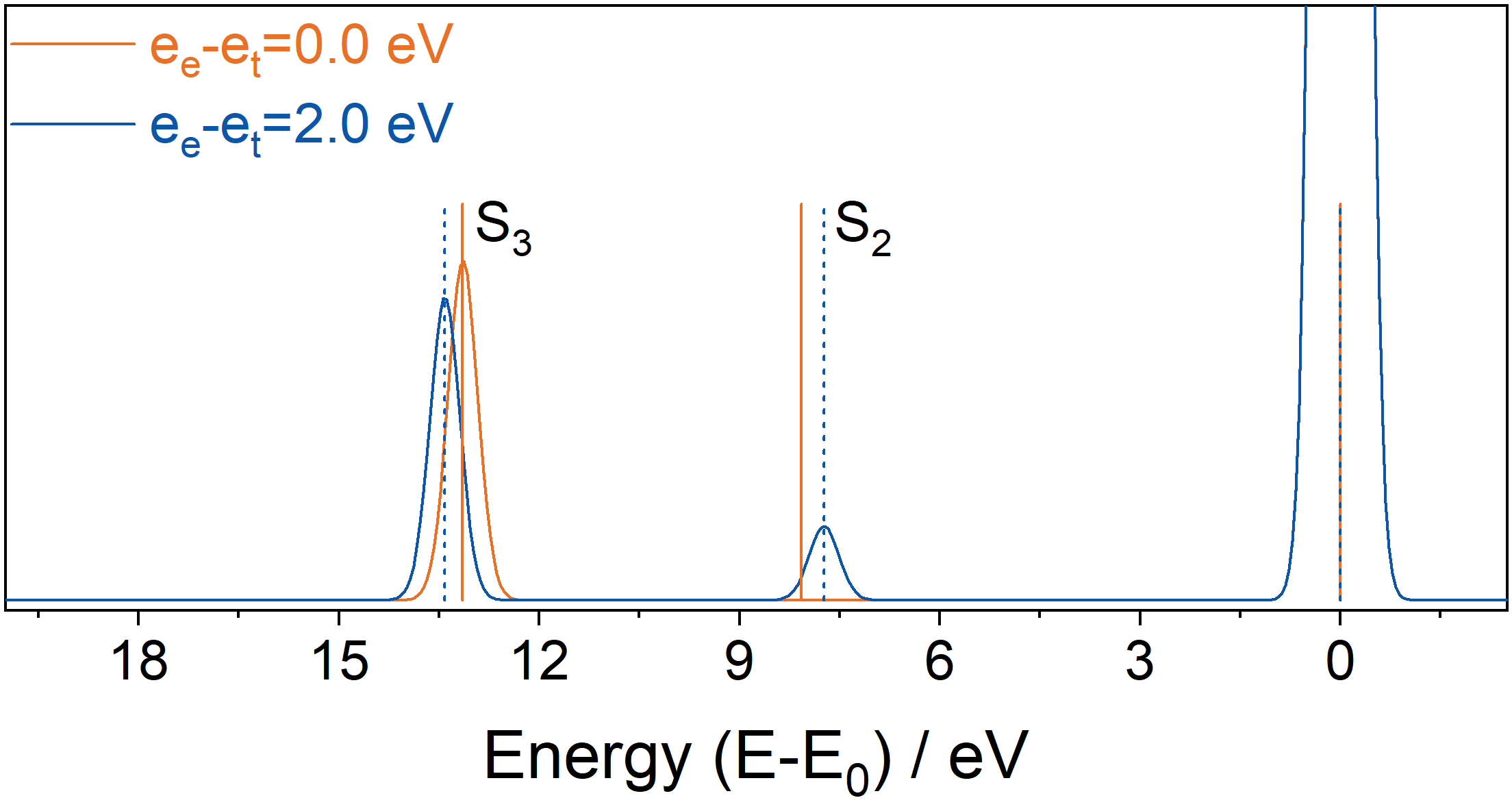}
    \caption{The simulated x-ray photoelectron spectroscopy (XPS) spectra of the toy model (red). The spectrum with $e_e=e_t$, i.e.,~the crystal field splitting is zero (blue). The spectrum with $e_e-e_t=1.97$~eV. The vertical lines indicate the energies of the final states in the toy model for the two cases. When $e_e=e_t$, the nonbonding state is present at the energy of $8.07$~eV, but its spectral intensity is zero.}
    \label{fig:toy}
\end{figure}

When $e_e=e_t \equiv e$, i.e.~the crystal field splitting is zero, the nonbonding state $|n\rangle$ is fully decoupled from the $|0\rangle$ and $|b\rangle$ states, and the Hamiltonian becomes

\begin{equation}
H=
\begin{pmatrix}
  0   & \sqrt{t_e^2+t_t^2} & 0  \\
  \sqrt{t_e^2+t_t^2} & \dfrac{et_e^2+e t_t^2}{t_e^2+t_t^2} & 0   \\
  0 & 0   & \dfrac{e t_t^2+e t_e^2}{t_e^2+t_t^2}      \notag
\end{pmatrix}.
\end{equation}

In this limit, the 1$s$ spectrum is composed of two states:~the bonding state and the antibonding states (of the $|0\rangle$ and $|b\rangle$ states), which produce the 0 and 13~eV ($S_3$) peaks in the experimental data, respectively. When $e_e\neq e_t$ in the realistic case with the crystal field splitting, contribution of the nonbonding state $|n\rangle$ shows up in the spectrum in between the two peaks, yielding the $S_2$ satellite. Since the coupling of the $|b\rangle$ and $|n\rangle$ states is in general very weak compared with the hybridization strength between the $|b\rangle$ and $|0\rangle$ states ($=\sqrt{t_e^2+t_t^2}$), it can be viewed as a weak perturbation to the bonding and antibonding formation of the Ti--O network. Thus, it does not yield a large peak shift nor intensity modulation to the two peaks, as observed in Fig.~\ref{fig:cluster}.

Figure~\ref{fig:toy} shows the XPS spectra computed from the toy model. To simulate the studied Ti oxides, model parameters are extracted from the cluster model studied above as  $v_e=r\sqrt{4V^2_{eg}}$, $v_t=r\sqrt{6V^2_{t2g}}$, $e_e=\Delta_{\rm CT}+6Dq$, and $e_t=\Delta_{\rm CT}-4Dq$, where $V_{eg}=4.01$~eV, $V_{t2g}=-2.33$~eV, charge-transfer energy $\Delta_{\rm CT}=3.00$~eV, and the crystal field splitting $10Dq=1.97$~eV. To consider the effect of the higher-order Ti--O charge-transfer processes, the hopping parameter values are rescaled by a constant factor $r=0.65$. The toy model reproduces the spectra of the cluster model with many-body electronic configurations in Fig.~\ref{fig:cluster} nicely. The vertical lines indicate the eigenstate energies of the final state Hamiltonian. When the crystal field splitting is absent, the nonbonding state is present between 0~eV and $S_3$ peaks, but not visible in the spectrum. With the finite crystal field splitting, the nonbonding state appears at $\sim$7~eV with a smaller spectral intensity than the other two peaks.

It is worth noting that, in the cluster model result of Fig.~\ref{fig:cluster}(b), the nonbonding satellite $S_2$ has a nonzero spectral intensity even when $10Dq=0.00$~eV unless the hopping parameter for the Ti $e_g$ and $t_{2g}$ orbitals is isotropic ($V_{eg}=V_{t2g}$). 
However, the visibility of the $S_2$ satellite in the toy model seems to concern only the presence of the crystal field splitting. This is an artifact of this simple toy model. The $S_2$ satellite gets a finite intensity for $V_{eg}\neq V_{t2g}$ by including higher-order states with two Ti $d$ electrons (and two ligand holes) to the toy model above. Only when fully isotropic ($10Dq$ is zero and $V_{eg}=V_{t2g}$), see Fig.~\ref{fig:cluster}(a), the nonbonding satellite $S_2$ cannot be excited in the XPS process. 

\begin{table}
    \centering
        \caption{The parameter values adopted in the TiO$_6$ cluster model for simulating \ce{SrTiO3} in electronvolts.}
    {\tabcolsep =0.3cm
    \renewcommand\arraystretch{1.4}
    \begin{tabular}{lccccc}
    \hline \hline
         & $U_{dd}$ & $U_{cd}$ & $V_{eg}$ & $V_{t_{2g}}$ & $10Dq$  \\
    \hline
          & 4.50 & 5.40  & 4.01  & -2.33  & 1.97   \\
    \hline \hline
    \end{tabular}
    }
    \label{tab:para}
\end{table}

\begin{table}
    \centering
     \caption{The hopping parameter values adopted in the TiO$_6$ cluster model for simulating \ce{TiO2} in electronvolts.}
    {\tabcolsep =0.3cm
    \renewcommand\arraystretch{1.4}
    \begin{tabular}{lccccc}
    \hline \hline
         & $V_{B_{1g}}$ & $V_{A_g}$ & $V_{A_g'}$ & $V_{B_{2g}}$ & $V_{B_{3g}}$  \\
    \hline
          & 3.87 & 3.85  & -2.35 & -2.15  & -2.25   \\
    \hline \hline
    \end{tabular}
    }
    \label{tab:hop}
\end{table}

\begin{table}
     \vspace{-0.5cm}
    \centering
        \caption{The configuration diagonal energies in the TiO$_6$ cluster model for \ce{SrTiO3} and \ce{TiO2} in electronvolts.}
    {\tabcolsep =0.3cm
   \renewcommand\arraystretch{1.4}
    \begin{tabular}{ccccccc}
    \hline \hline
        & $|d^0\rangle$ & 0 & 0.0 & \\
    & $|d^1$$\underline{L}^1\rangle$ & $\Delta$ & 3.0 & \\
    & $|d^2$$\underline{L}^2\rangle$ & 2$\Delta$+$U_{dd}$ & 10.5  &\\
   & $|d^3$$\underline{L}^3\rangle$ & 3$\Delta$+3$U_{dd}$ & 22.5 &\\
   & $|d^4$$\underline{L}^4\rangle$ & 4$\Delta$+6$U_{dd}$ & 39.0  &\\
   & $|d^5$$\underline{L}^5\rangle$ & 5$\Delta$+10$U_{dd}$ & 60.0  &\\
   & $|d^6$$\underline{L}^6\rangle$ & 6$\Delta$+15$U_{dd}$ & 85.5 & \\
    \hline \hline
    \vspace{+0.2cm}
    \end{tabular}
    }
    \label{tab:eng}
\end{table}

\section{Parameters of the TiO$_6$ cluster model}

The parameters defining the cluster model of SrTiO$_3$ are summarized in Table~\ref{tab:para}. 
The electron hopping amplitude $V_{eg}$ ($V_{t2g}$) of the Ti $e_g$ ($t_{2g}$) orbital with the nearest-neighboring molecular orbital of the ligands and the crystal field splitting $10Dq$ are read from the tight-binding model constructed from the LDA bands. The interaction parameters $U_{dd}$, $U_{cd}$ are set to the same values as in the LDA+DMFT AIM. 
The charge-transfer energy $\Delta_{\rm CT}$ is set to $3.0$~eV.
The Ti 1$s$ core level spectra are rather insensitive to the $\Delta_{\rm CT}$ value in a realistic range.
Table~\ref{tab:hop} shows the hopping amplitude of different orbitals [$B_{1g}(xy)$, $A_g(3z^2-r^2)$, $A_g'(x^2-y^2)$, $B_{2g}(zx)$, and  $B_{3g}(yz)$] in TiO$_2$.
Table~\ref{tab:eng} summarizes the configuration diagonal energies accounting for the interaction $U_{dd}$ and the charge-transfer energy $\Delta_{\rm CT}$ up to $|d^6\underline{L}^6\rangle$ configurations.

\bibliographystyle{apsrev4-1}
\bibliography{main}

\end{document}


\title{Satellites in the Ti~1$s$ core level spectra of \ce{SrTiO3} and \ce{TiO2} 
\newline Supplementary Information}

\author{Atsushi~Hariki}
\author{Keisuke~Higashi}
\author{Tatsuya Yamaguchi}
\affiliation{Department of Physics and Electronics, Graduate School of Engineering, Osaka Prefecture University 1-1 Gakuen-cho, Nakaku, Sakai, Osaka 599-8531, Japan.}

\author{Jiebin~Li}
\author{Curran~Kalha}
\affiliation{Department of Chemistry, University College London, 20 Gordon Street, London WC1H 0AJ, United Kingdom.}%

\author{Manfred~Mascheck}
\affiliation{Scienta Omicron GmbH, Limburger Strasse 75, 65232 Taunusstein, Germany.}

\author{Susanna~K.~Eriksson}
\author{Tomas Wiell}
\affiliation{Scienta Omicron AB, P.O. Box 15120, 750 15 Uppsala, Sweden.}

\author{Frank~M.~F.~de~Groot}
\affiliation{Inorganic Chemistry \& Catalysis, Debye Institute for Nanomaterials Science, Utrecht University, Universiteitsweg 99, 3584 CG, Utrecht, The Netherlands.}

\author{Anna~Regoutz}
\email{a.regoutz@ucl.ac.uk}
\affiliation{Department of Chemistry, University College London, 20 Gordon Street, London WC1H 0AJ, United Kingdom.}%

\date{\today}

\maketitle

\subsection*{A.~X-ray Photoelectron Spectra}

\begin{figure*}[h]
    \includegraphics[width=0.7\textwidth]{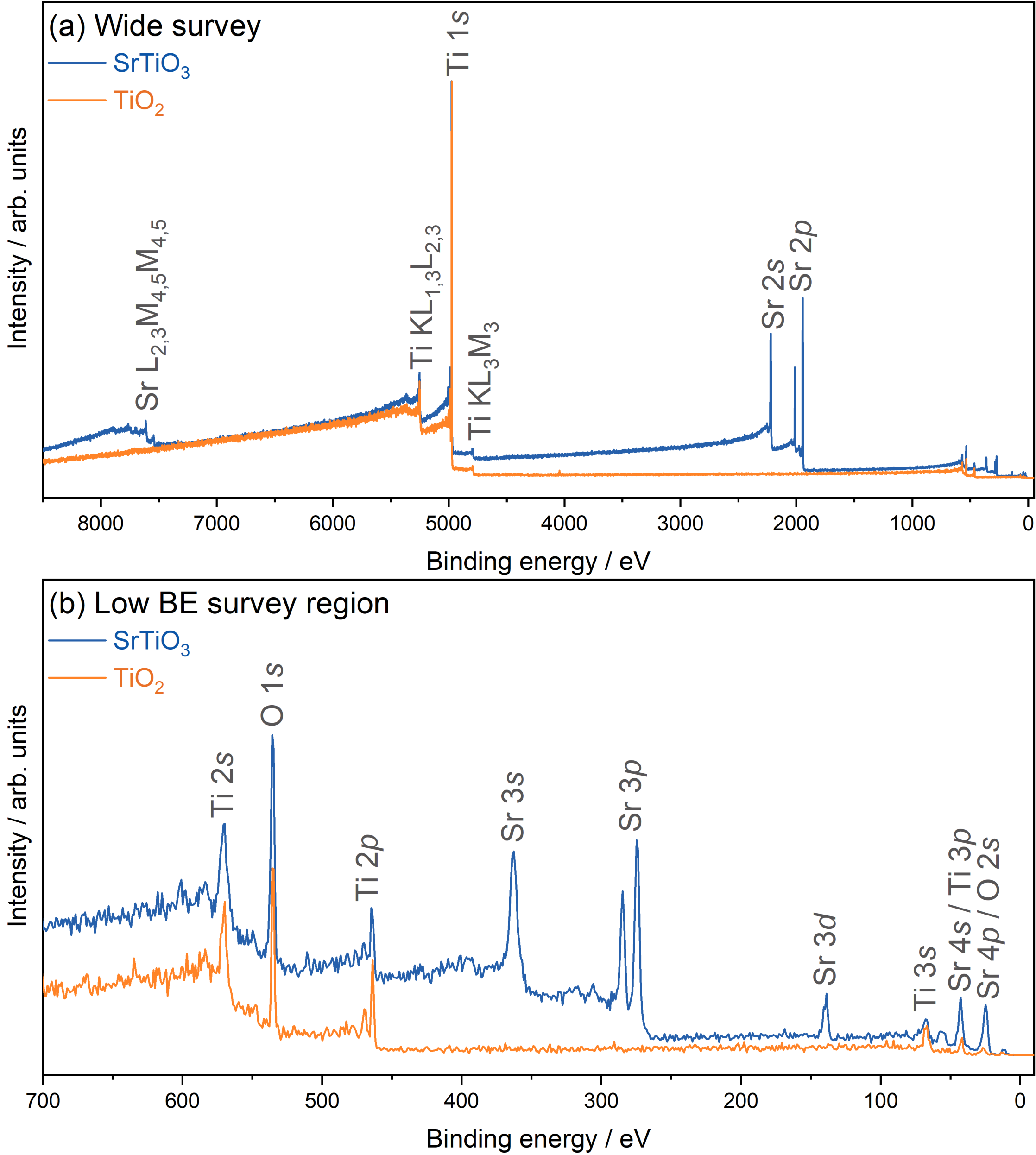}
    \caption{HAXPES survey spectra of \ce{SrTiO3} and (b) \ce{TiO2}, including (a) wide HAXPES BE range and (b) low BE range. All major core and Auger lines are indicated.}
    \label{fig:PES_survey}
\end{figure*}

\begin{figure*}[h]
    \includegraphics[width=0.7\textwidth]{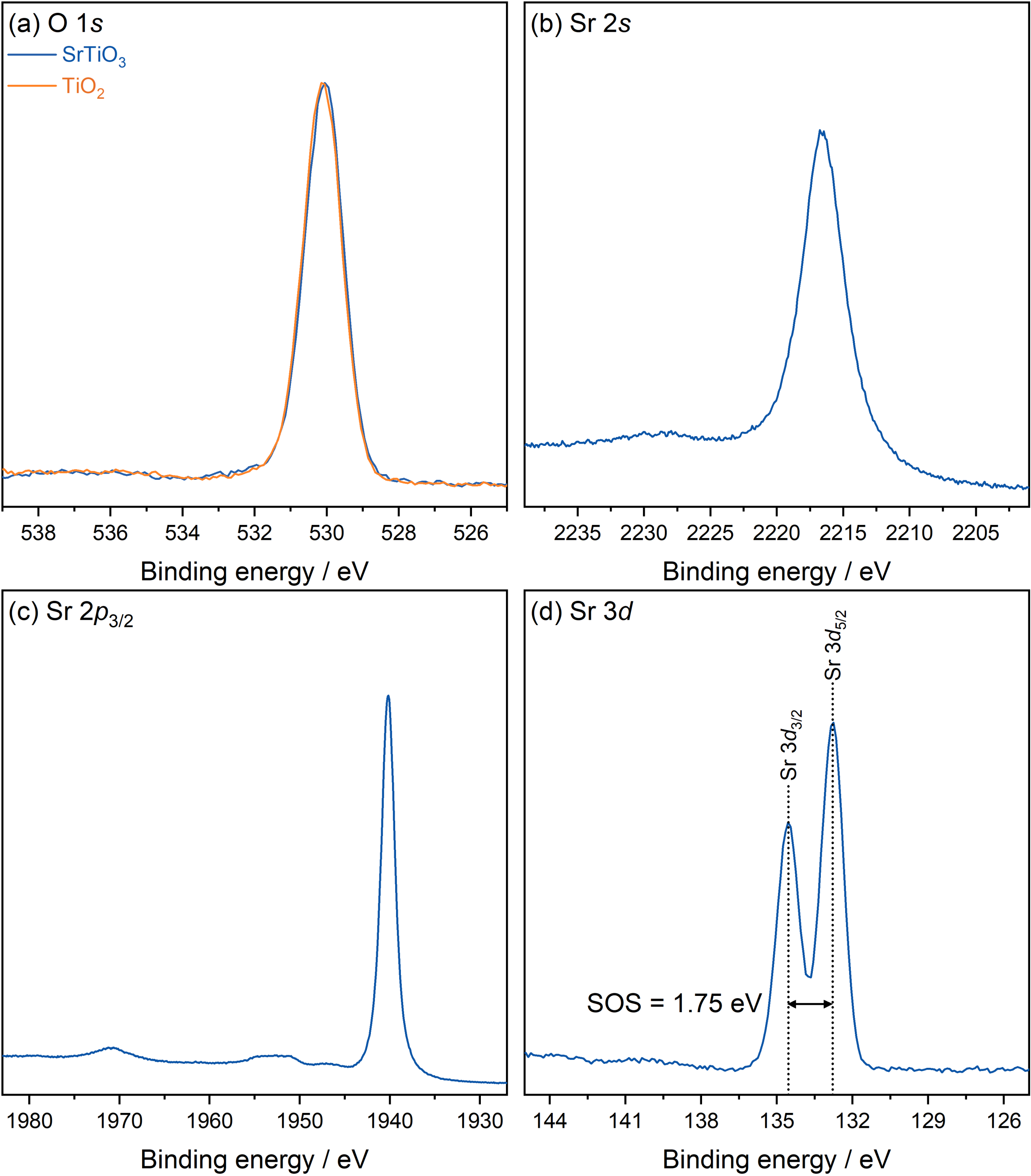}
    \caption{HAXPES core level spectra of \ce{SrTiO3} and (b) \ce{TiO2}, including (a) O~1$s$, (b) Sr~2$s$, (c) Sr~2$p_{3/2}$ and (d) Sr~3$d$. In (d) the spin orbit splitting (SOS) of Sr~3$d$ is also included.}
    \label{fig:PES_CLs}
\end{figure*}

\begin{figure*}[h]
    \includegraphics[width=0.5\textwidth]{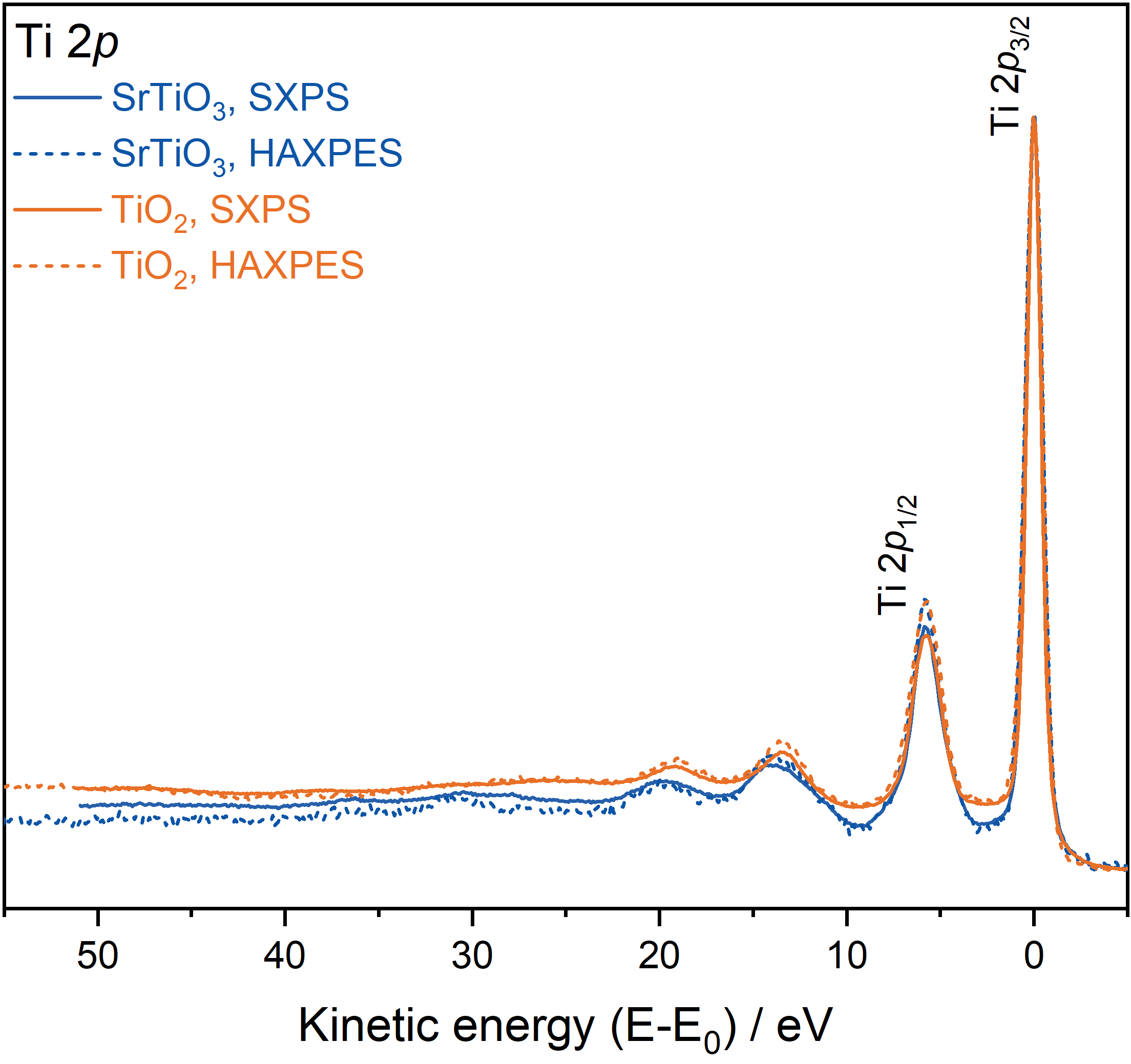}
    \caption{SXPS and HAXPES Ti~$2p$ core level spectra of \ce{SrTiO3} and (b) \ce{TiO2}.}
    \label{fig:Ti2p_SXHS}
\end{figure*}

\clearpage
\subsection*{B.~Double counting dependence of the LDA+DMFT result}

\begin{figure*}[h]
    \includegraphics[width=0.90\textwidth]{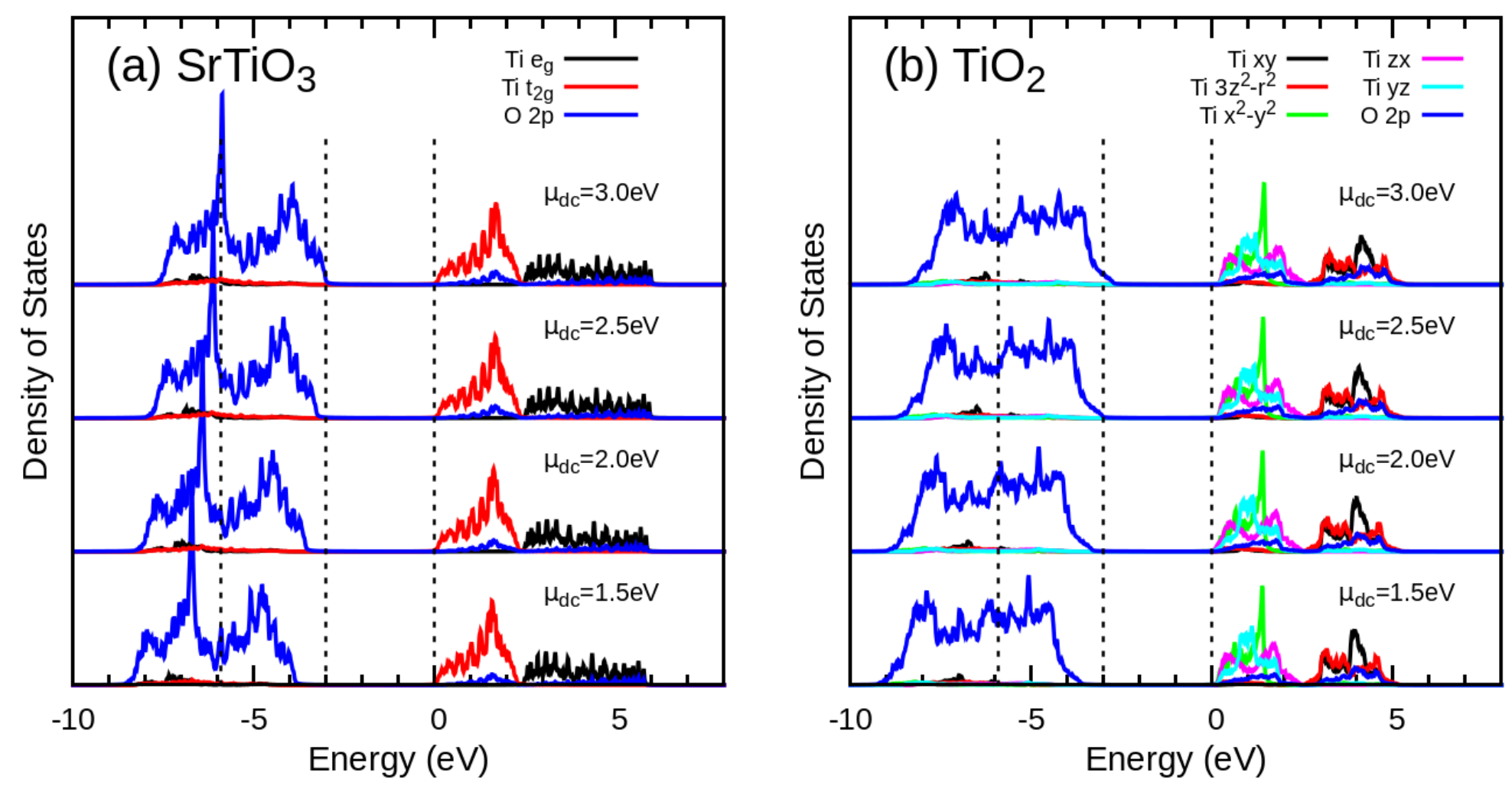}
    \caption{The LDA+DMFT PDOS of (a) \ce{SrTiO3} and (b) \ce{TiO2} calculated for different double-counting correction values $\mu_{\rm dc}$.}
    \label{fig:ti_dos}
\end{figure*}

Figure~\ref{fig:ti_dos} shows the LDA+DMFT projected density of states (PDOS) calculated for different double-counting correction $\mu_{\rm dc}$ values.
 $\mu_{\rm dc}$ accounts for the electron-electron interaction implicitly present in the LDA result.
 The Ti 3$d$ bare energy $\varepsilon_d^{\rm bare}$ is given by subtracting the $\mu_{dc}$ value from the LDA energy $\varepsilon_d^{\rm LDA}$, i.e.~$\varepsilon_d^{\rm bare}=\varepsilon_d^{\rm LDA}-\mu_{\rm dc}$.
 Practically, $\mu_{dc}$ renormalises the energy splitting between the Ti 3$d$ and O 2$p$ states, see Figure~\ref{fig:ti_dos}. 
Thus, for the studied tetravalent Ti oxides with a band gap between filled $d$ and empty $p$ bands, the appropriated $\mu_{\rm dc}$ value can be determined by comparing with the experimental band gap, and valence and inverse photoemission spectra.

\begin{figure}[h]
    \includegraphics[width=0.50\textwidth]{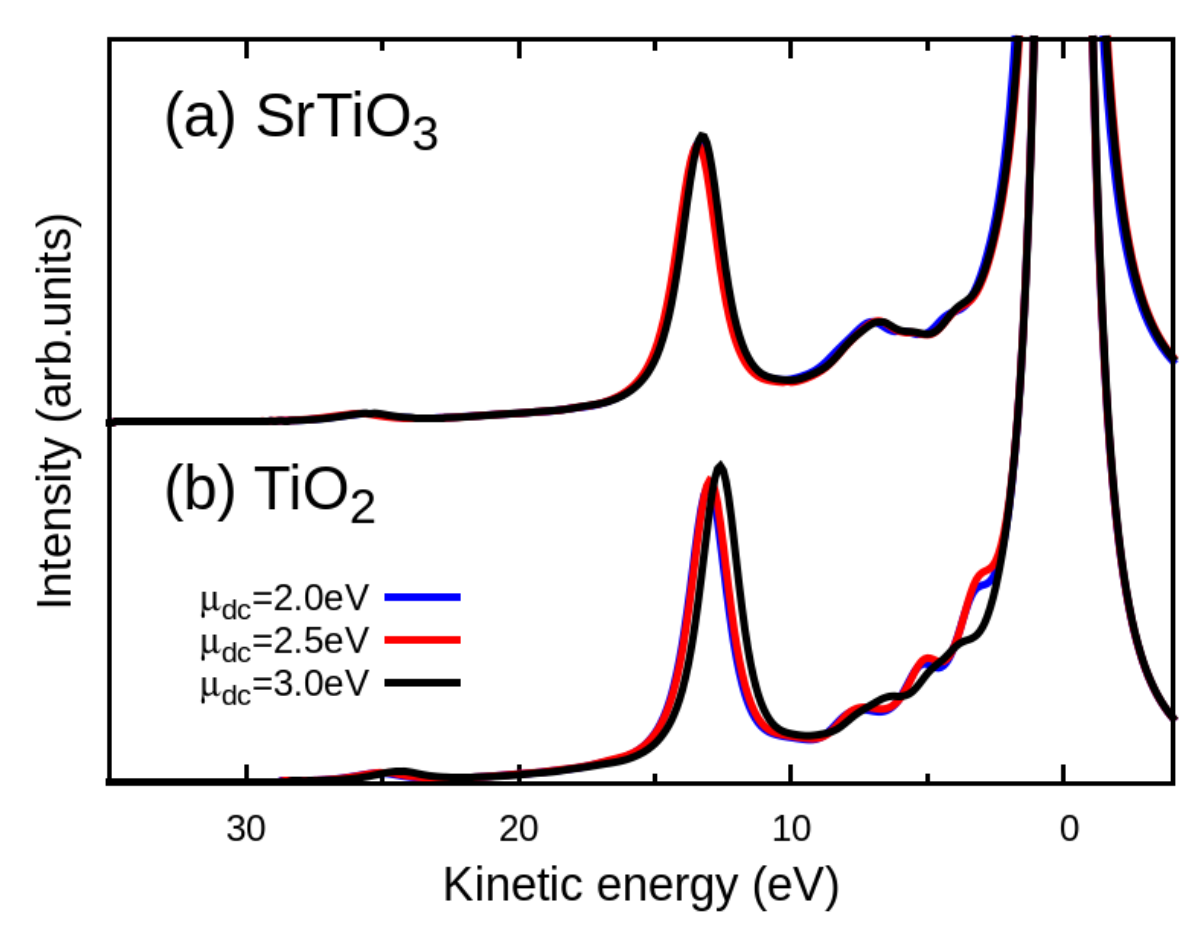}
    \caption{Ti~1$s$ HAXPES spectra of (a) \ce{SrTiO3} and (b) \ce{TiO2} calculated by the LDA+DMFT method for different double-counting correction $\mu_{\rm dc}$ values.}
    \label{fig:ti_xps}
\end{figure}

For both compounds, $\mu_{\rm dc}=3.0$~eV reproduces well the experimental band gap and the valence photoemission (see Figure~2 in the main manuscript) and reported inverse photoemission data.~\cite{Tezuka94,Ohtomo04} 
The 1$s$ core-level spectra are rather insensitive to the used $\mu_{\rm dc}$ value, see Figure~\ref{fig:ti_xps}.
This is expected because the binding energies of the satellites are determined predominantly by the Ti--O hybridization strength, and thus a small change to the Ti--O energy splitting is a weak perturbation to the satellites.

\begin{figure*}[h]
    \includegraphics[width=0.5\textwidth]{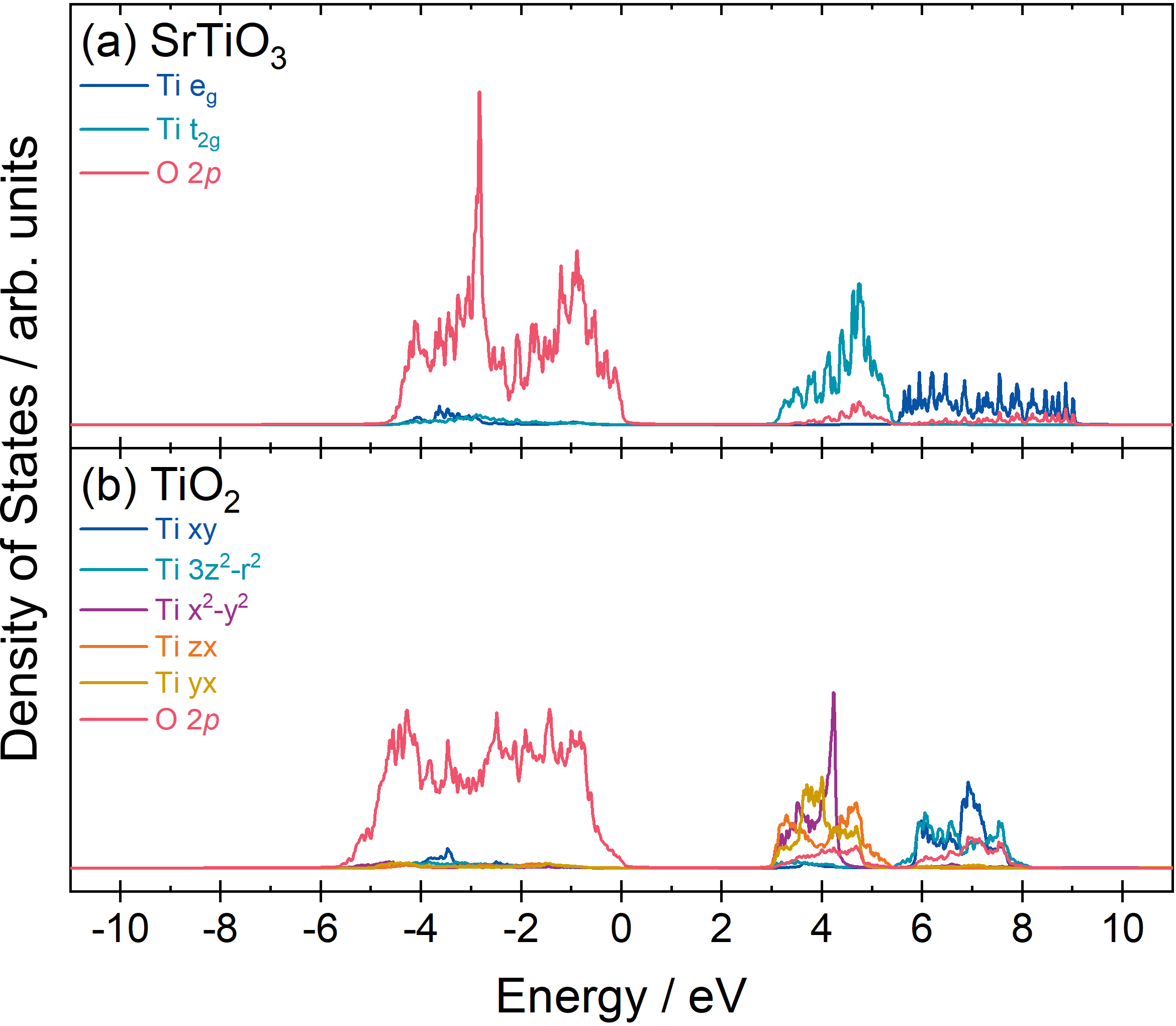}
    \caption{LDA+DMFT projected density of states (PDOS) for (a) \ce{SrTiO3} and (b) \ce{TiO2}. $\mu_{\rm dc}=3.0$~eV was used.}
    \label{fig:PDOS_raw}
\end{figure*}

\newpage
\bibliography{si}